\documentclass[12pt]{article}
\usepackage{epsfig}
\usepackage{amssymb,amsmath}
\usepackage{setspace}
\pdfoutput=1
\usepackage{subfigure}
\usepackage{amssymb,amsmath}
\usepackage{graphicx}
\usepackage{color}
\usepackage{cancel}
\usepackage[colorlinks=true
,urlcolor=blue
,citecolor=blue
,linkcolor=blue
,pagecolor=blue
,linktocpage=true
,pdfproducer=medialab
]{hyperref}
\def\bi{\begin{itemize}}
\def\ei{\end{itemize}}
\newcommand{\Z}{{\mathbb Z}}

\def\alt{\lesssim}
\def\agt{\gtrsim}

\newcommand\prd[3]{{\it Phys.\ Rev.\ }{\bf D #1} (#2) #3}
\newcommand\prl[3]{{\it Phys.\ Rev.\ Lett.\ }{\bf #1} (#2) #3}

\usepackage{amsmath}

\newcommand{\bea}{\begin{eqnarray}}
\newcommand{\eea}{\end{eqnarray}}

\newcommand{\beq}{\begin{equation}}
\newcommand{\eeq}{\end{equation}}

 \makeatletter
\def\alt{\mathrel{\mathpalette\gl@align<}}
\def\agt{\mathrel{\mathpalette\gl@align>}}
\def\gl@align#1#2{\lower.6ex\vbox{\baselineskip\z@skip\lineskip\z@
\ialign{$\m@th#1\hfil##\hfil$\crcr#2\crcr\sim\crcr}}} \makeatother

\usepackage{longtable}

\begin{document}
%

\begin{center}
{
\bf\LARGE
Revisit a realistic intersecting D6-brane with modified soft SUSY terms}
\\[10mm]
{\large Imtiaz~Khan$^{\,\star\,\heartsuit}$} \footnote{E-mail: \texttt{ikhan@itp.ac.cn}},
{\large Waqas~Ahmed$^{\,\diamond}$} \footnote{E-mail: \texttt{waqasmit@hbpu.edu.cn}},
{\large Tianjun~Li$^{\,\star\,\heartsuit}$} \footnote{E-mail: \texttt{tli@itp.ac.cn}},\\[1mm]
{\large Shabbar~Raza$^{\,\ast}$} \footnote{E-mail: \texttt{shabbar.raza@fuuast.edu.pk}}
\\[10mm]
\centerline{$^{\star}$ \it
CAS Key Laboratory of Theoretical Physics,}
\centerline{\it
Institute of Theoretical Physics, Chinese Academy of Sciences,}
\centerline{\it Beijing 100190, China}
\vspace*{0.2cm}
\centerline{ $^\heartsuit $\it 
School of Physical Sciences, University of Chinese Academy of Sciences, No. 19A Yuquan Road,} 
\centerline{\it  Beijing 100049, China}
\vspace*{0.2cm}
\centerline{$^{\diamond}$ \it Center for Fundamental Physics and School of Mathematics and Physics, Hubei Polytechnic University,}
\centerline{\it  Huangshi 435003,China}
\vspace*{0.2cm}
\centerline{$^{\ast}$ \it
Department of Physics, Federal Urdu University of Arts, Science and Technology,}
\centerline{\it Karachi 75300, Pakistan}

\vspace{1.cm} 
{\bf Abstract}
\end{center}
Because there are a few typos in the supersymmetry breaking sfermion masses and trilinear soft term, regarding the current Large Hadron Collider (LHC) and dark matter searches, we revisit a three-family Pati-Salam model based on intersecting D6-branes in Type IIA string theory on a $\mathbf{T^6/(\Z_2\times \Z_2)}$ orientifold with a realistic phenomenology. We study the viable parameter space and discuss the spectrum consistent with the current LHC Supersymmetry searches along with the dark matter relic density bounds from the Planck 2018 data. For the gluinos and first two generations of sfermions, we observe that the gluino mass is in the range [2, 14] TeV, the squarks mass range is [2, 13] TeV and the sleptons mass is in the range [1, 5] TeV. We achieve the cold dark matter relic density consistent with 5$\sigma$ Planck 2018 bounds via A-funnel and coannihilation channels such as stop-neutralino, stau-neutralino, and chargino-neutralino. Except for a few chargino-neutralino coannihilation solutions, these solutions also satisfy current nucleon-neutralino spin-independent and spin-dependent scattering cross-sections and may be probed by future dark matter searches.


\thispagestyle{empty}

\newpage

\addtocounter{page}{-1}

\baselineskip 18pt

\section{Introduction}
\label{Intro}

In the Large Hadron Collider (LHC), no evidence has been found for physics beyond the Standard Model (SM). The observation of the Higgs boson with mass around $m_{h}=$125 TeV \cite{Aad:2012tfa,CMS}, whose properties are in good agreement with the SM predictions, put challenges in the extension of SM which has been proposed to provide gauge coupling unification of the electromagnetic, weak and strong interactions \cite{gaugeunification}, a natural explanation to the hierarchy between the electroweak symmetry breaking (EWSB) and Plank scale. One of the promising candidates to deal with the hierarchy problem is supersymmetry (SUSY). The observed Higgs boson mass and null results from SUSY searches of the ATLAS and CMS collaboration have put the low-energy SUSY under stress. Thus, one can think to find comparatively natural and non-minimal solutions. Besides the hierarchy problem, the promising motivation for physics at the LHC accessible scale is to explain the observed dark matter (DM) in terms of relic particle density produced in the thermal freeze-out mechanism in the early universe. In the supersymmetric standard models (SSMs) with conserved R-parity, the Lightest Supersymmetric Particle (LSP) is stable and can be the dark matter candidate \cite{Jungman:1995df}. According to the recent searches gluino mass $m_{\tilde{g}}\gtrsim 2.2$ TeV for the first two generation squark mass $m_{\tilde{q}} \gtrsim 2$ TeV ~\cite{ATLAS:2017mjy, Vami:2019slp, CMS:2017okm}.  In literature, several interesting scenarios have been recently discussed, particularly the one called "Super-Natural SUSY"~\cite{Leggett:2014hha, Du:2015una}. In this framework, in the Minimal Supersymmetric Standard Model (MSSM), no residual fine tuning is left in the presence of no-scale supergravity boundary conditions ~\cite{Cremmer:1983bf} and Giudice-Masiero (GM) mechanism~\cite{Giudice:1988yz} despite a relatively heavy spectrum.

String theory is one of the most promising candidates for quantum gravity. Thus, string phenomenology aims to construct SM or SSMs from the string theory with moduli stabilization and without chiral exotics and try to make unique predictions that can be probed in LHC and other future experiments. In this article, we are interested in updating the phenomenological study of the intersecting D-brane models~\cite{Berkooz:1996km,
Ibanez:2001nd, Blumenhagen:2001te, CSU, Cvetic:2002pj, Cvetic:2004ui, Cvetic:2004nk, Cvetic:2005bn, Chen:2005ab, Chen:2005mj, Blumenhagen:2005mu}. For the intersecting D-brane model building, the realistic SM fermion Yukawa couplings can be realized only within the Pati-Salam gauge group \cite{PS}. Three-family Pati-Salam models have been constructed systematically in Type IIA string theory on the $\mathbf{T^6/(\Z_2\times \Z_2)}$ orientifold with intersecting D6-branes~\cite{Cvetic:2004ui}, and was found that one model has a realistic phenomenology: the tree-level gauge coupling unification is realized naturally around the string scale, the Pati-Salam gauge symmetry can be broken down to the SM close to the string scale, the small number of extra chiral exotic states can be decoupled via the Higgs mechanism and strong dynamics, the SM fermion masses and mixing can be accounted for, the low-energy sparticle spectra may potentially be tested at the LHC, and the observed dark matter relic density may be generated for the lightest neutralino as the LSP, and so on~\cite{Chen:2007px, Chen:2007zu, Li:2014xqa}. In short, as far as we know,  this is one of the best globally consistent string models that is phenomenologically viable from the string scale to the EWSB scale.

Because there are a few typos in the supersymmetry breaking sfermion masses and trilinear soft term, the purpose of this study is to highlight the 
differences in parameter space associated with the soft SUSY-breaking terms in our previous work \cite{Li:2014xqa} and recalculated in \cite{Sabir:2022hko} with $\mu >$0. In this work, we display the viable parameter space satisfying the collider and DM bounds along with the Higgs mass bounds.
We show that in our present scans, we have A/H-resonance solutions, chargino-neutralino coannihilation, stau-neutralino coannihilation, and stop-neutralino coannihilation. In the case of resonance solutions, $m_{A/H}$ is about 2 TeV or so.  In the case of chargino-neutralino coannihilation is concerned the NLSP chargino mass can be between 0.7 TeV to 2.3 TeV and the NLSP stau is in the mass range of 0.2 TeV to 1.8 TeV. As far as the NLSP stop solutions are concerned we we have solutions from 0.15 TeV to 0.9 TeV. Most of the parameter space related to this scenario has already been probed by the LHC SUSY searches. It should also be noted that the above-mentioned solutions, except for some of the chargino-neutralino solutions, are consistent with the ongoing and future astrophysical dark matter experiments. 

The paper is organized as follows. In section (\ref{model}) we highlight the model's features related to our study. In section (\ref{constraints}) we review the detail of the range of values we employed and the phenomenological constraints we impose. We discuss the numerical results of our scanning in section (\ref{Discussion}). Section (\ref{summary}) gives a summary and conclusion.

\section{The realistic Pati-Salam model from the intersecting D6-branes compatified on a $\mathbf{T^6/(\Z_2\times \Z_2)}$ orientifold }
\label{model}
We are going to focus on the realistic intersecting D6-brane model ~\cite{Cvetic:2004ui} with modified soft SUSY terms calculated in \cite{Sabir:2022hko}. Ignoring the CP-violating phase, the SSB terms by nonzero F-terms of the dilaton $F^S$ and three complex structure moduli $F^{U^i}$, where $i = 1,2,3$ can be parametrized by the $\Theta_{1}$, $\Theta_{2}$, $\Theta_{3}$, $\Theta_{4}$ and the gravitino mass $m_{3/2}$. Here $\Theta_{4} \equiv \Theta_{S}$ for the dilaton case. The relationship among the $\Theta$'s is given as~\cite{Li:2014xqa}
\begin{align}
	\sum_{i=1}^4\Theta^2_i=1.
	\label{Theta4}
\end{align}
 The SSB terms at the grand unification (GUT) scale in terms of these parameters can be written as~\cite{Sabir:2022hko} 
\begin{align}
M_1 & =m_{3/2} (0.519615 \Theta _1+0.34641 \Theta _2+0.866025 \Theta _3) ,\nonumber\\
M_2 & =m_{3/2} (0.866025 \Theta _2-0.866025 \Theta _4) ,\nonumber\\
M_3 & =m_{3/2} ( 0.866025 \Theta _2+0.866025 \Theta _3) ,\nonumber\\
A_0 &= m_{3/2}(-0.292797 \Theta_1-1.43925 \Theta_2-0.573228 \Theta_3 +0.573228 \Theta_4) ,\nonumber\\
 \widetilde{m}^2_{L}& = m_{3/2}{}^2 \Big(1-2.02977 \Theta_1{}^2+0.75 \Theta_1 \Theta_2-1.5 \Theta_1 \Theta_4-0.0440466 \Theta_2{}^2-1.5 \Theta_2
\Theta_3 \nonumber\\
&\quad\quad\quad +0.286907 \Theta_3{}^2+0.75 \Theta_3 \Theta_4+0.286907 \Theta_4{}^2\Big) ,\nonumber\\
\widetilde{m}^2_{R}& = m_{3/2}{}^2 \Big(1-0.0880932 \Theta_1{}^2-1.5 \Theta_1 \Theta_2+0.75 \Theta_1 \Theta_3+0.75 \Theta_1 \Theta_4-0.0880932
\Theta_2{}^2\nonumber\\
&\quad\quad\quad+0.75 \Theta_2 \Theta_3+0.75 \Theta_2 \Theta_4-0.419047 \Theta_3{}^2-1.5 \Theta_3 \Theta_4-2.40477 \Theta_4{}^2\Big)
~,~\nonumber\\
\widetilde{m}_{H_u}^2&=\widetilde{m}_{H_d}^2=m_{3/2}^2(1.0-(1.5 \Theta_{3}^2)-(1.5 \Theta_{4}^2)) .
\label{ssb}
\end{align}
All the above results are subject to the constraint in eq.(\ref{Theta4}).
Here, $M_{1,2,3}$ are the gauginos masses for the gauge groups $U(1)_Y$, $SU(2)_L$, $SU(3)_c$ respectively, $A_0$ is a common trilinear scalar coupling term and $\widetilde{m}_L$ and $\widetilde{m}_R$ are the soft mass terms for the left-handed and right-handed squarks and sleptons respectively, and $\widetilde{m}_{H_{u,d}}$ are the SSB Higgs soft mass terms. The gauginos and Higgs soft masses are the same as in the case \cite{Li:2014xqa}. The trilinear coupling $A0$ equation is different only by the coefficients of $\Theta$'s with no new extra terms, unlike the case of $\widetilde{m}_L{}^2$ and $\widetilde{m}_R{}^2$. In the left-handed squarks soft mass square term $\widetilde{m}_L{}^2$ in eq.(\ref{ssb}), apart from the coefficients of $\Theta$'s we have some additional terms like, $\Theta_{1}\Theta_{4}$ and $\Theta_{2}\Theta_{3}$. Similarly, the right-handed sleptons soft mass square term $\widetilde{m}_R{}^2$ irrespective of the coefficients of $\Theta$'s, we also have some additional new terms like, $\Theta_{1}\Theta_{3}$, $\Theta_{1}\Theta_{4}$, $\Theta_{2}\Theta_{3}$, and $\Theta_{2}\Theta_{4}$. These terms predict that our parameters space differs from the previously discussed results\cite{Li:2014xqa}, and the details are discussed in section \ref{Discussion}.

\section{Scanning procedure and phenomenological constraints}
\label{constraints}
We employ the ISAJET~7.85 package~\cite{ISAJET} to perform random scans over the parameter space of the presented above intersecting D6-brane model.
Following \cite{Li:2014xqa}, we can parametrize the three independent $\Theta_i$ with $i = 1, 2, 3$ that enter the soft masses in (\ref{ssb}) in terms of $\gamma_{1}$, $\gamma_{2}$, and $\Theta_{4}$ as, 
\begin{align}
	&\Theta_1 = \cos(\beta)\cos(\alpha)\sqrt{1-\Theta_{4}^{2}} ,~\nonumber\\
	&\Theta_2 = \cos(\beta)\sin(\alpha)\sqrt{1-\Theta_{4}^{2}} ,~\nonumber \\
	&\Theta_3 = \sin(\beta)\sqrt{1-\Theta_{4}^{2}} ,~\nonumber\\
	&{\rm where}\quad \alpha  \equiv 2 \pi \gamma_{1} ,~ \beta \equiv 2 \pi \gamma_{2} .
	\label{Theta's}
\end{align}
We perform random scans over the following ranges of the model parameters:
\begin{align}
	0\leq & \gamma_1  \leq 1  ~,~\nonumber \\
	0\leq & \gamma_2  \leq 1 ~,~\nonumber \\
	0\leq &  \Theta_4  \leq 1 ~,~\nonumber \\
	0\leq & m_{3/2}  \leq 15 ~ \rm{TeV} ~,~\nonumber \\
	2\leq & \tan\beta  \leq 60~,
	\label{input_param_range}
\end{align}
where tan$\beta$ is the ratio of vacuum expectation values (VEVs) of the Higgs fields.
We use the $m_t = 173.3$ GeV \cite{:2009ec}. We employ the Metropolis-Hastings algorithm as described in \cite{Belanger:2009ti, Baer:2008ksh}. We have done our scans with $\mu>0$ and collected the data points that satisfy the requirement of a successful radiative EWSB (REWSB). Besides, we have also selected those points with the lightest neutralino being the LSP. After collecting the data, we impose the following constraints that the LEP2 experiment set on charged sparticles masses  \cite{Patrignani:2016xqp}
\begin{eqnarray} 
	\label{LEP2}
	m_{\tilde t_1},m_{\tilde b_1} \gtrsim 100 \; {\rm GeV} ~,~\\
	m_{\tilde \tau_1} \gtrsim 105 \; {\rm GeV}  ~,~\\
	m_{\tilde \chi_{1}^{\pm}} \gtrsim 103 \; {\rm GeV}~,~
\end{eqnarray}
and the combined Higgs mass reported by the ATLAS and CMS collaborations \cite{Khachatryan:2016vau}
\begin{align}\label{eqn:mh}
	m_{h} = 125.09 \pm 0.21(\rm stat.) \pm 0.11(\rm syst.)~GeV .
\end{align}  
Because of the theoretical uncertainty in the calculation of $m_h$,
we consider the following range for the Higgs mass \cite{Slavich:2020zjv, Allanach:2004rh}
\begin{align}\label{eqn:higgsMassLHC}
	122~ {\rm GeV} \leq m_h \leq 128~ {\rm GeV}. 
\end{align}
Furthermore, we use the IsaTools package \cite{bsg,bmm} to implement the following observables B-physics constraints\cite{CMS:2014xfa, Amhis:2014hma}:
\begin{align}\label{eqn:Bphysics}
	1.6\times 10^{-9} \leq ~ {\rm BR}(B_s \rightarrow \mu^+ \mu^-) ~
	\leq 4.2 \times10^{-9} \quad,\\ 
	2.99 \times 10^{-4} \leq  ~ {\rm BR}(b \rightarrow s \gamma) ~
	\leq 3.87 \times 10^{-4} \quad,\\
	0.70\times 10^{-4} \leq ~ {\rm BR}(B_u\rightarrow\tau \nu_{\tau})~
	\leq 1.5 \times 10^{-4} \quad.
\end{align}

In addition to the above constraints, we consider the following conditions on the gluino and first/second generations squarks masses from the LHC and Planck bound based on \cite{ATLAS:2017mjy, Vami:2019slp, CMS:2017okm, Planck:2018nkj}
\begin{align}
m_{\tilde{g}} \gtrsim 2.2 \, {\rm TeV}\ ({\rm for}\  m_{\tilde{q}}\gtrsim 2)\, {\rm TeV} \nonumber\\
	0.114 \leq \Omega_{\rm \tilde{\chi}_1^0}h^2 ({\rm Planck}) \leq 0.126  
	\end{align} 
	
\section{Numerical results and discussion}
\label{Discussion}

\begin{figure}[h!]
	\centering \includegraphics[width=7.90cm]{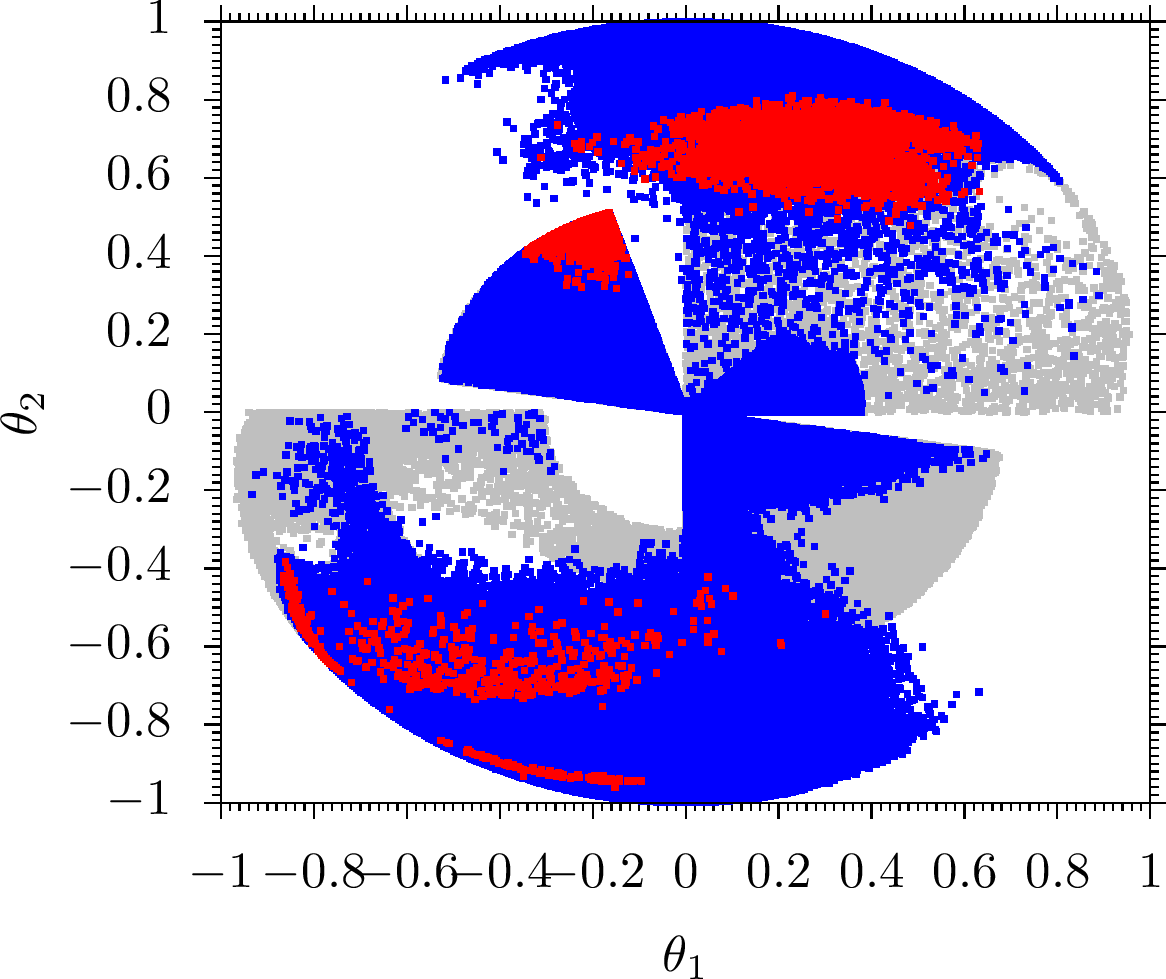}
	\centering \includegraphics[width=7.90cm]{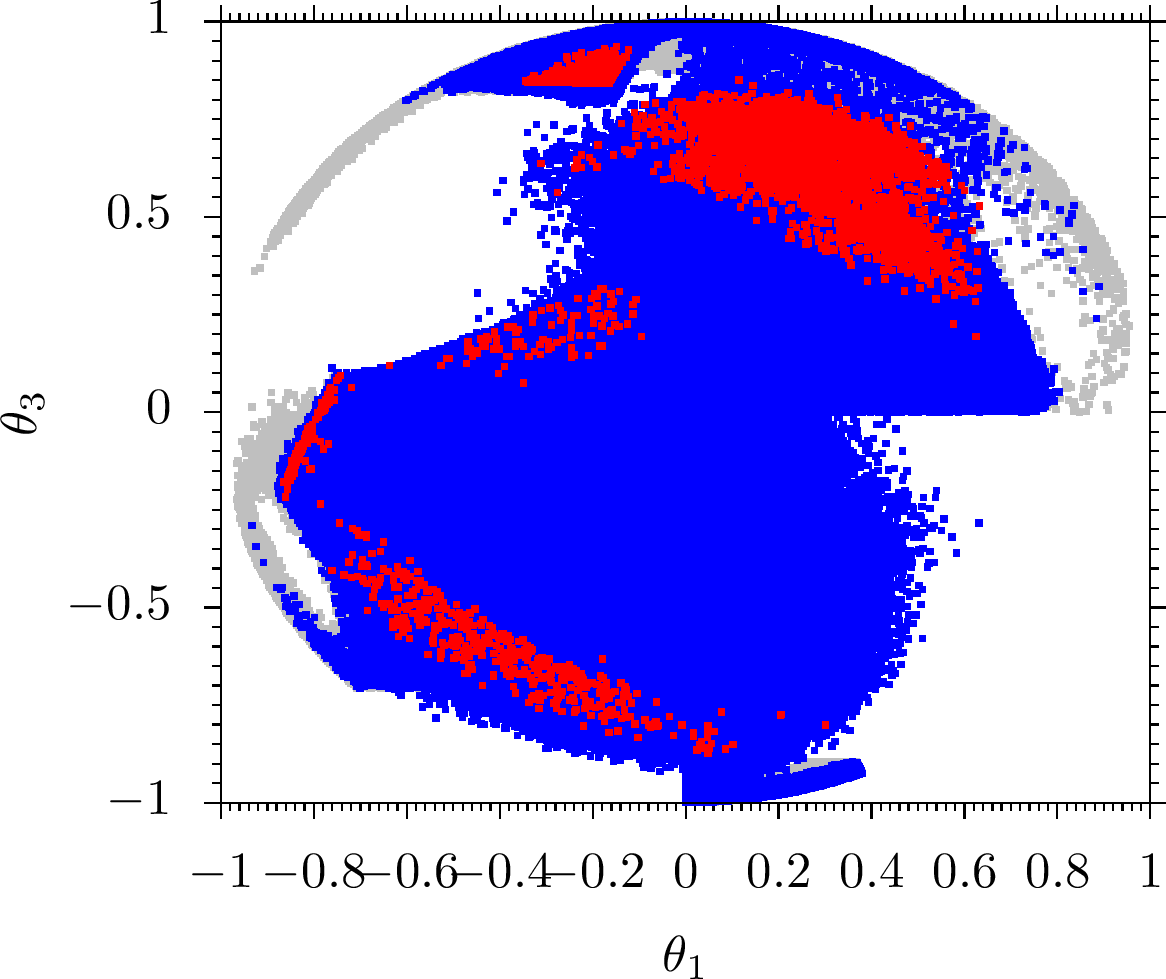}
	\centering \includegraphics[width=7.90cm]{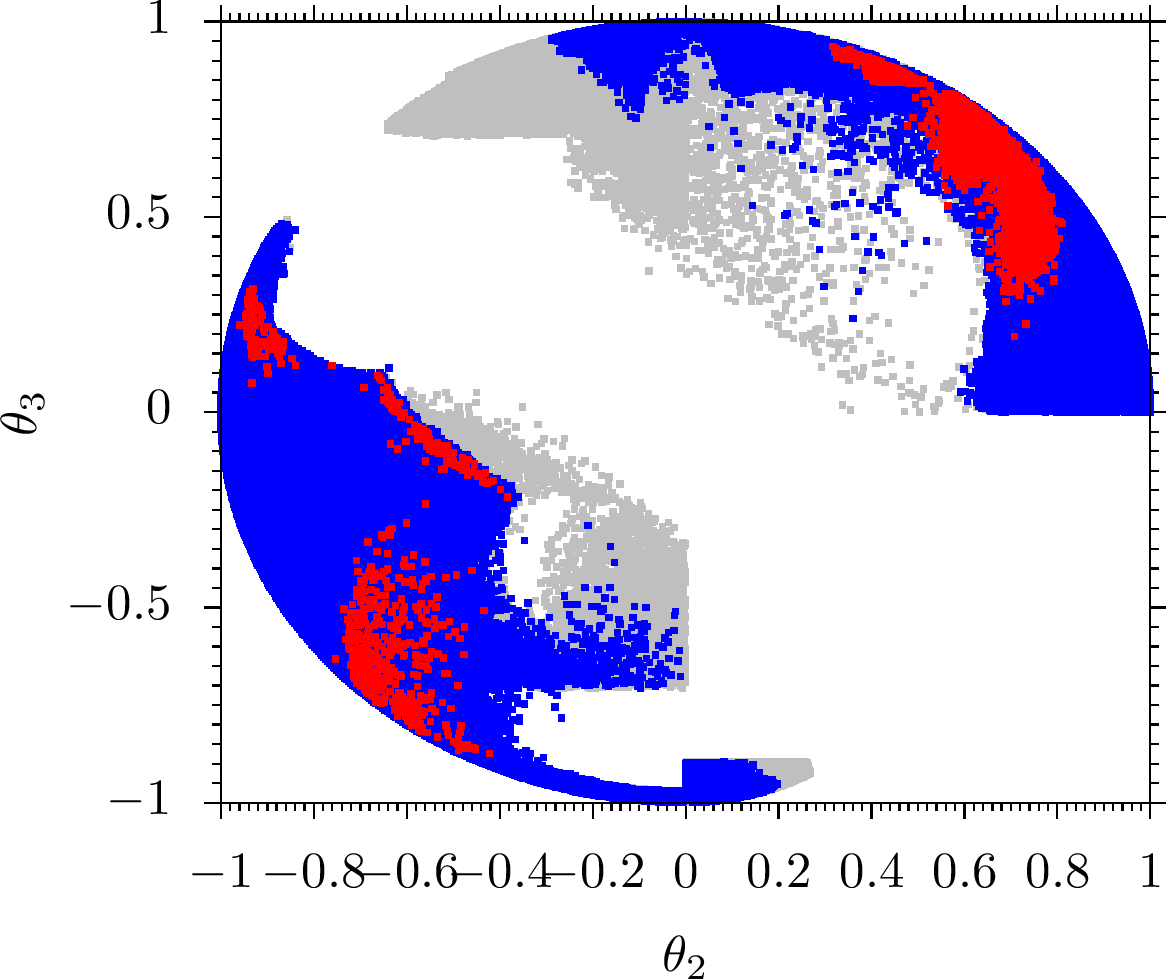}
	\caption{ Grey points satisfy the REWSB and yield LSP neutralino.  Blue points satisfy LEP, Higgs mass bound, B-physics, and LHC sparticle mass bounds.  Red points are a subset of blue points that satisfy 5 $\sigma$ Planck relic density bounds.}
	\label{1}
\end{figure}

In Fig. (\ref{1}), we show graphs for various parameters in eq. (\ref{Theta's}). We consider $\mu > 0$ and the color coding is as follows. Grey points satisfy the REWSB and yield LSP neutralino. Blue points satisfy LEP, Higgs mass bound, B-physics, and LHC sparticle mass bounds. Red points form a subset of blue points and satisfy Planck 2018 bounds on cold dark matter relic density within 5$\sigma$.

In our scanning,  we see in $\Theta_{1}$-$\Theta_{2}$ plane, the range of the red points for $\Theta_{1}$ is $-0.9\lesssim \Theta_1 \lesssim 0.7$, but most of the points are concentrated from $-0.4$ to $0.7$, while for $\Theta_{2}$ is $-0.9\lesssim \Theta_{2} \lesssim 0.8$. Also for the $\Theta_{2}$ most of the points are concentrated from $0.3$ to $0.8$ and $-0.8$ to $-0.4$. For the $\Theta_{1}$ and $\Theta_{2}$, we have red points solutions almost everywhere in the entire range except for the $\Theta_{2}$ where the solutions are in the $-0.4$ to $0.3$ range.  On the other hand, blue points are more or less everywhere in the plot. In $\Theta_{1}$-$\Theta_{3}$ plane, the concentration of red points favors the positive range as in the case of $\Theta_{1}$-$\Theta_{2}$ plane. We also see a small concentration of red points in the negative range for the small negative value of $\Theta_{1}$ and for the large negative value of $\Theta_{3}$. Blue points are almost everywhere in the plot in contrast to $\Theta_{1}$-$\Theta_{2}$ plane, as we have the density of points around the center of the plot. In the $\Theta_{3}$-$\Theta_{2}$ plane, here again, we see the concentration of red points favors the positive range for $\Theta_{2}$ and $\Theta_{3}$. But we also see a small concentration of red points in the negative range smaller than that of the positive range for a small negative value of $\Theta_{2}$ and a large negative value of $\Theta_{3}$. For all points, we see a polarization kind of pattern compared to other planes and having no points in the center of the plot. 
 
\begin{figure}[h!]
	\centering \includegraphics[width=7.90cm]{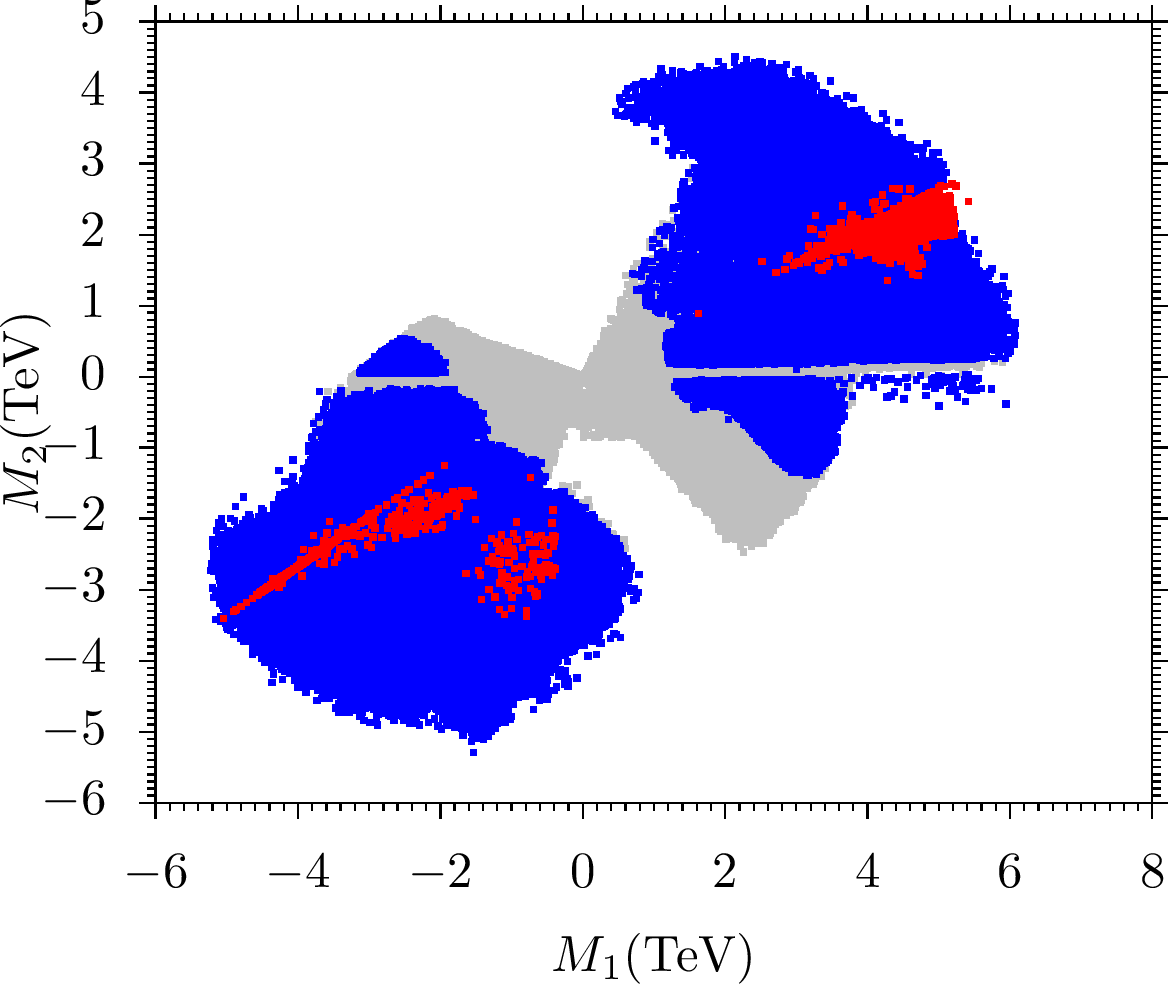}
	\centering \includegraphics[width=7.90cm]{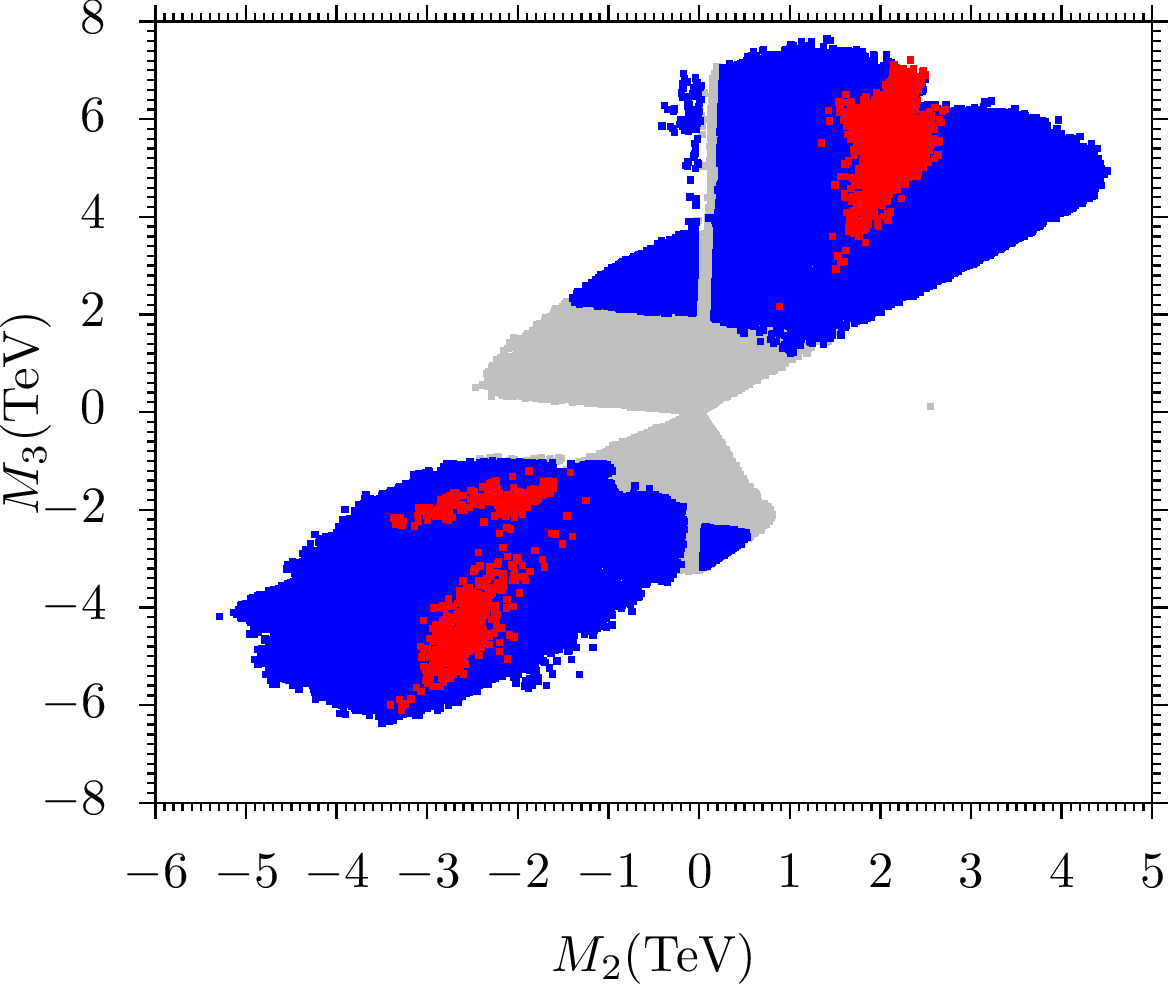}
	\centering \includegraphics[width=7.90cm]{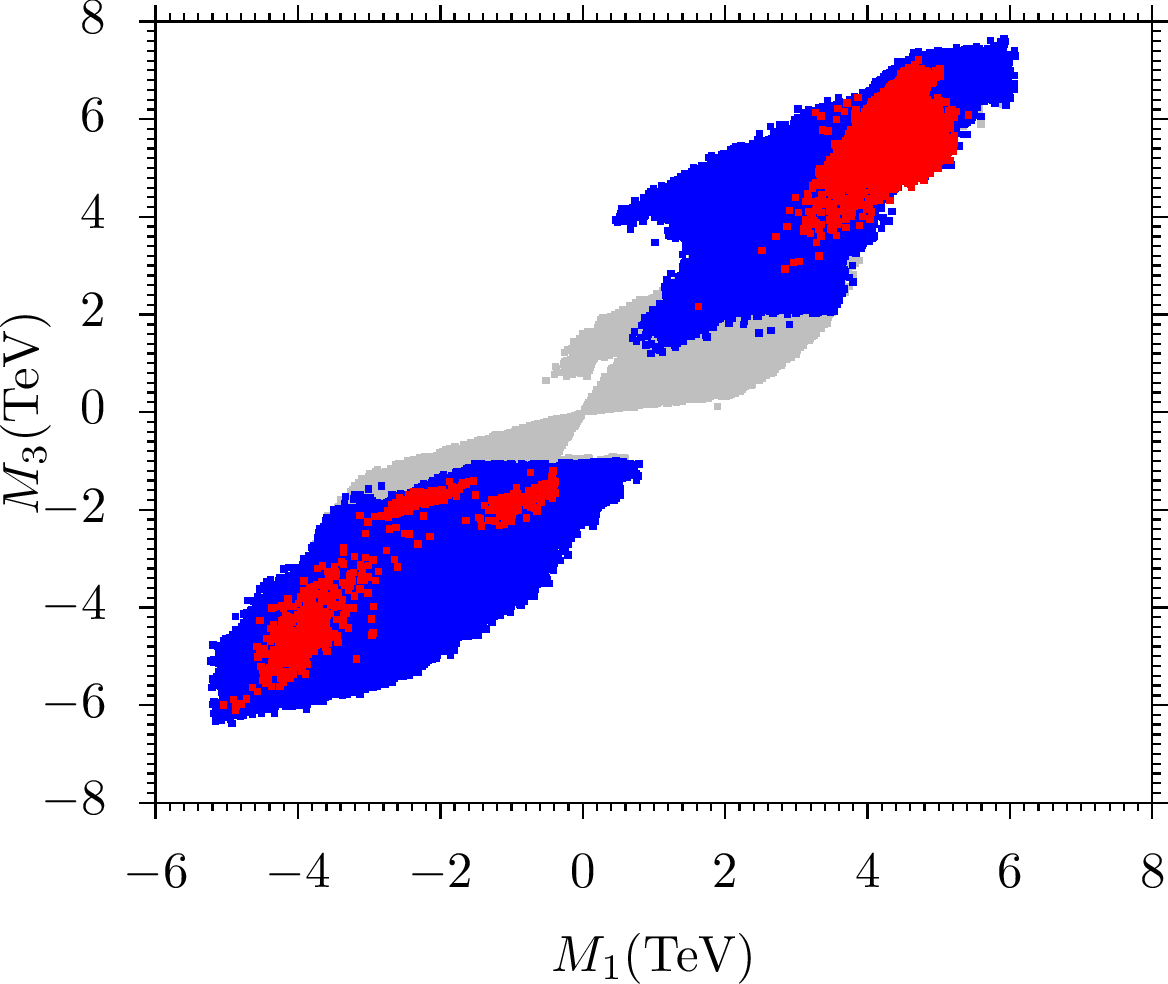}
	\caption{ Plots of results in $M_1$ - $M_2$, $M_1$ - $M_3$ and $M_3$ - $M_2$ planes. The color coding and the panel description are the same as in Fig. (\ref{1}).} 
	\label{2}
\end{figure} 

\begin{figure}[h!]
	\centering \includegraphics[width=7.90cm]{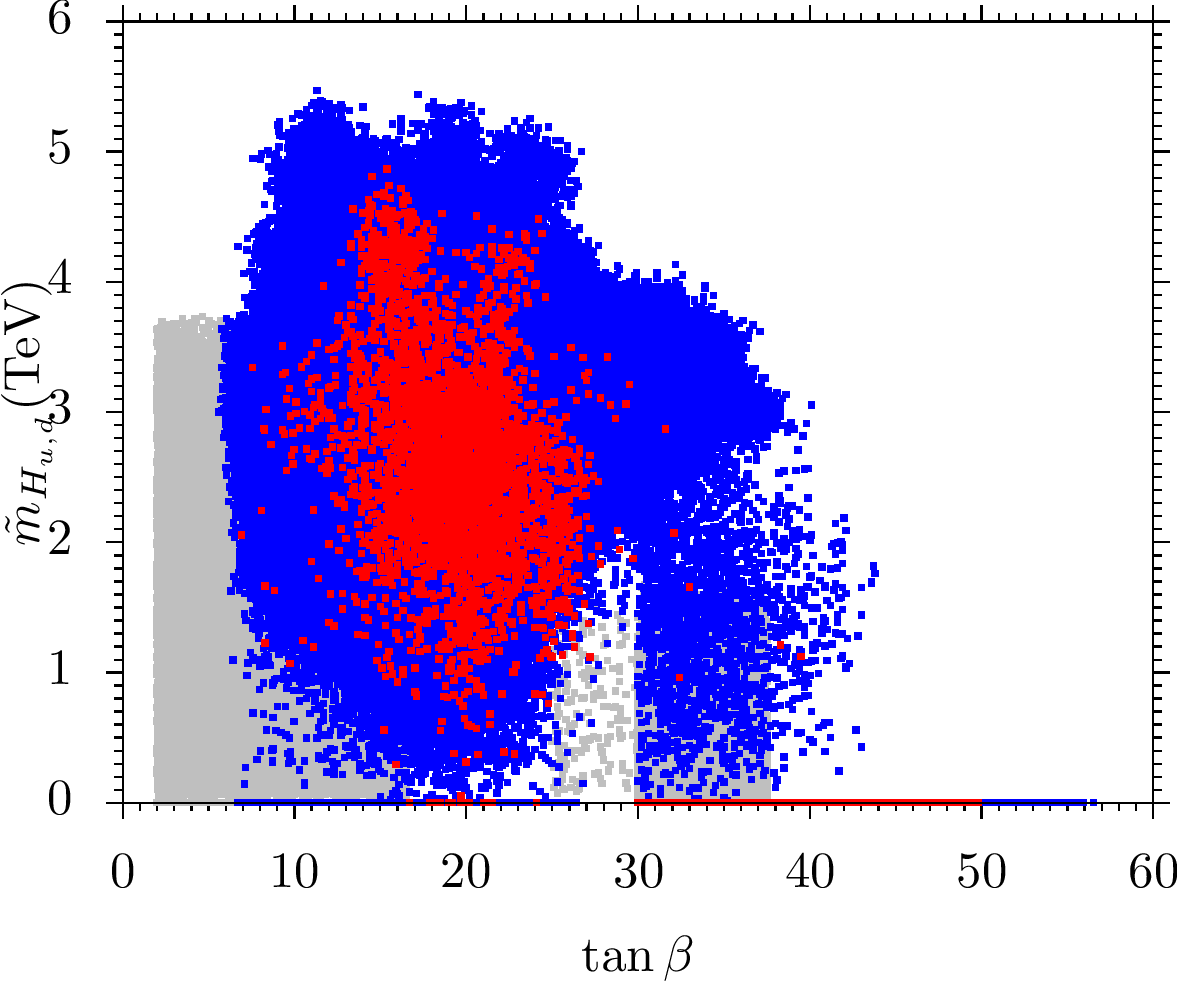}
	\centering \includegraphics[width=7.90cm]{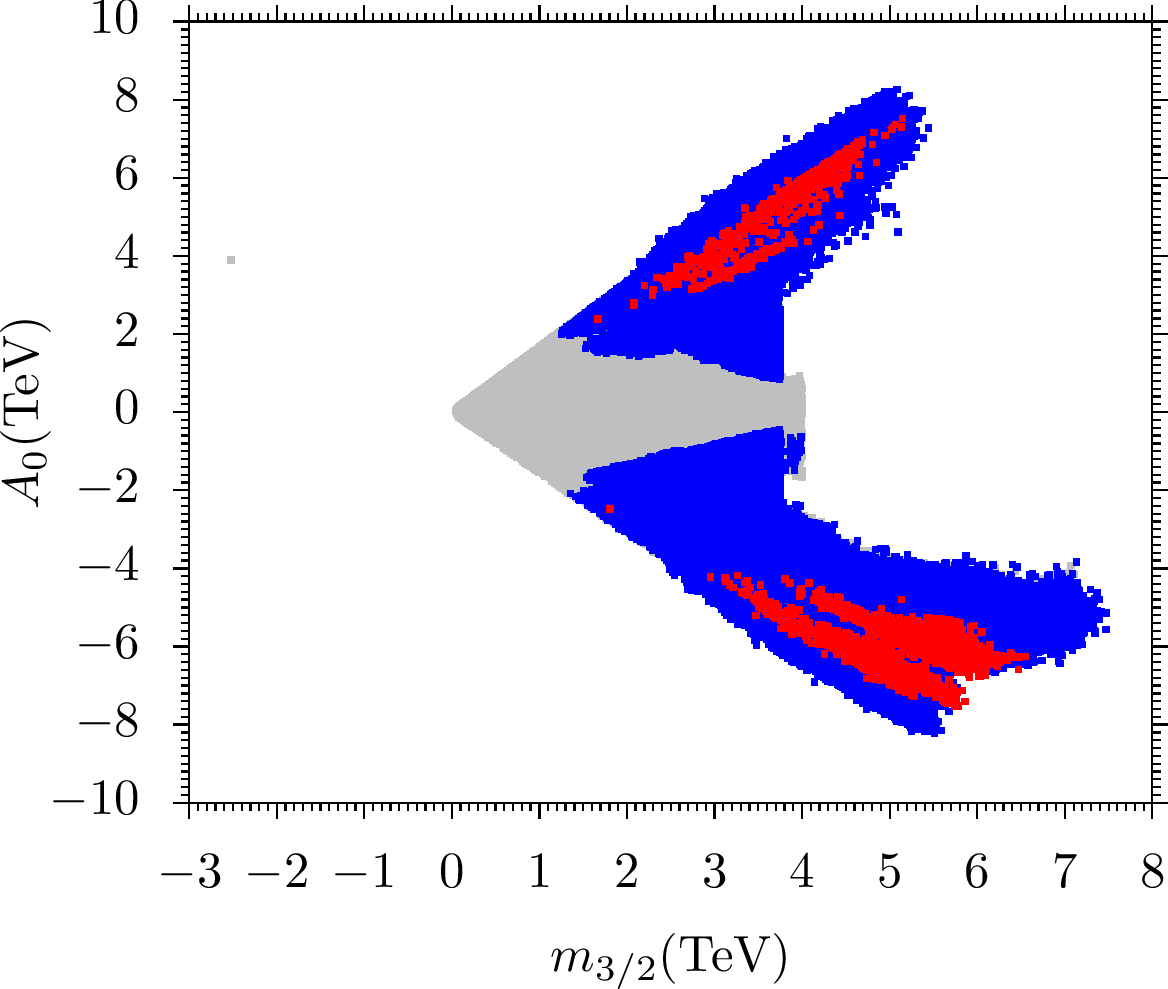}
	\centering \includegraphics[width=7.90cm]{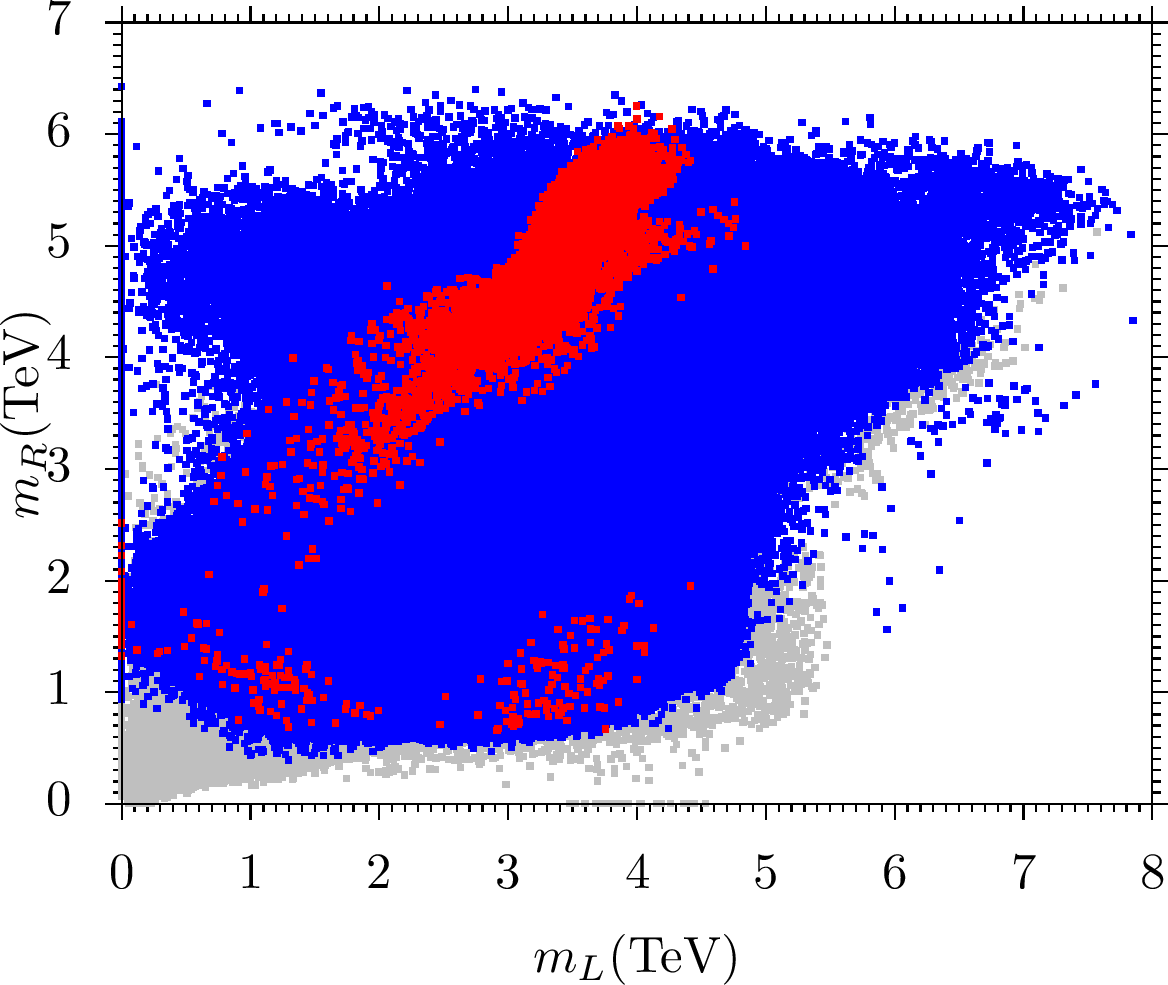}
	\caption{ Plots in $\tan\beta$ - $\widetilde{m}_{H_{u,d}}$, $A_0-m_{3/2}$, and $\widetilde{m}_L$ - $\widetilde{m}_R$ planes. The color coding and the panel description are the same as in Fig.(\ref{1}).}
	\label{3}
\end{figure}
  
 We calculate the (SSB) parameters given in Eqs. (\ref{ssb}). We present the results in Fig. (\ref{2}) and (\ref{3}), color coding is the same as that of Fig. (\ref{1}). We present the $M_1$ - $M_2$ plane in the top left panel. Red points are in the range from [-5, 6] TeV for $M_1$, but the large density favors the positive range from [4, 6] TeV. The density of red points smaller than that of the positive range was also concentrated for the negative values [-5, -2] TeV for $M_1$. For the blue points, we have solutions almost everywhere for $M_1$ from [-5, 6] TeV except around $1$ TeV. We also see the concentration of red points at [1, 3] TeV in the positive range and [-3.5,-1] TeV in the negative range for $M_2$. For the blue points, we have solutions almost everywhere from [-5, 4.5] TeV for $M_2$. In short, we see a polarisation kind of pattern for red and blue points having no points in the center. We present the $M_2$ - $M_3$ plane in the top right panel. We see almost a similar pattern as that of $M_1$ - $M_2$ plane but a little difference can be observed as, the red points are concentrated at [2, 3] TeV and [-3.5,-2] TeV for $M_2$, and at [4, 7] TeV and [-2, -6] TeV for $M_3$. For blue points, we have solutions for $M_2$ everywhere from [-5, 4.5] TeV, and for $M_3$, [-6, 7.5] TeV except in the middle. In short for all the points, again we see a similar pattern as that of the $M_1$ - $M_2$ plane. Finally, we see the $M_1$ - $M_3$ plot in the bottom panel. Similar to the two gaugino plots, here too we see a similar polarization pattern in solutions. The only difference is that since the ranges of $M_{1}$ and $M_{3}$ are relatively larger than $M_{2}$, this is why plot in $M_{1}-M_{3}$ looks slim.

 In Fig. (\ref{3}), we present the $\tan\beta$ - $\widetilde{m}_{H_{u,d}}$, $m_{3/2}$ - $A_0$, and $\widetilde{m}_L$ - $\widetilde{m}_R$ planes. Color coding is the same as in Fig. (\ref{1}). In the $\tan\beta-\widetilde{m}_{H_{u,d}}$ plane, we see that the red points solution are $7 \lesssim \tan\beta \lesssim 50$ with $0 \lesssim \widetilde{m}_{H_{u,d}} \lesssim 5$ TeV but most of the points concentrate in the range $\tan\beta=15$ to $\tan\beta=25$. For blue points, we have solution for $6.5 \lesssim \tan\beta \lesssim 57$ and $0 \lesssim \widetilde{m}_{H_{u,d}} \lesssim 5.4$ TeV. In the $m_{3/2}-A_0$ plane, most of the red points concentrate in the range [2, 7] TeV for $m_{3/2}$ and $|4|$ TeV to $|8|$ TeV for $A_{0}$. But one can see red solutions favor $A_{0}<0$.

 In the $\widetilde{m}_L-\widetilde{m}_R$ plane we see most of the concentrations of red points at [2, 4.5] TeV for $\widetilde{m}_L$ and at [3, 6] TeV for $\widetilde{m}_R$. For the blue points, we have solutions almost everywhere from [0, 7.9] TeV for $\widetilde{m}_L$ and from [0.5, 6.5] TeV for $\widetilde{m}_R$.
 
\begin{figure}[h!]
	\centering \includegraphics[width=7.90cm]{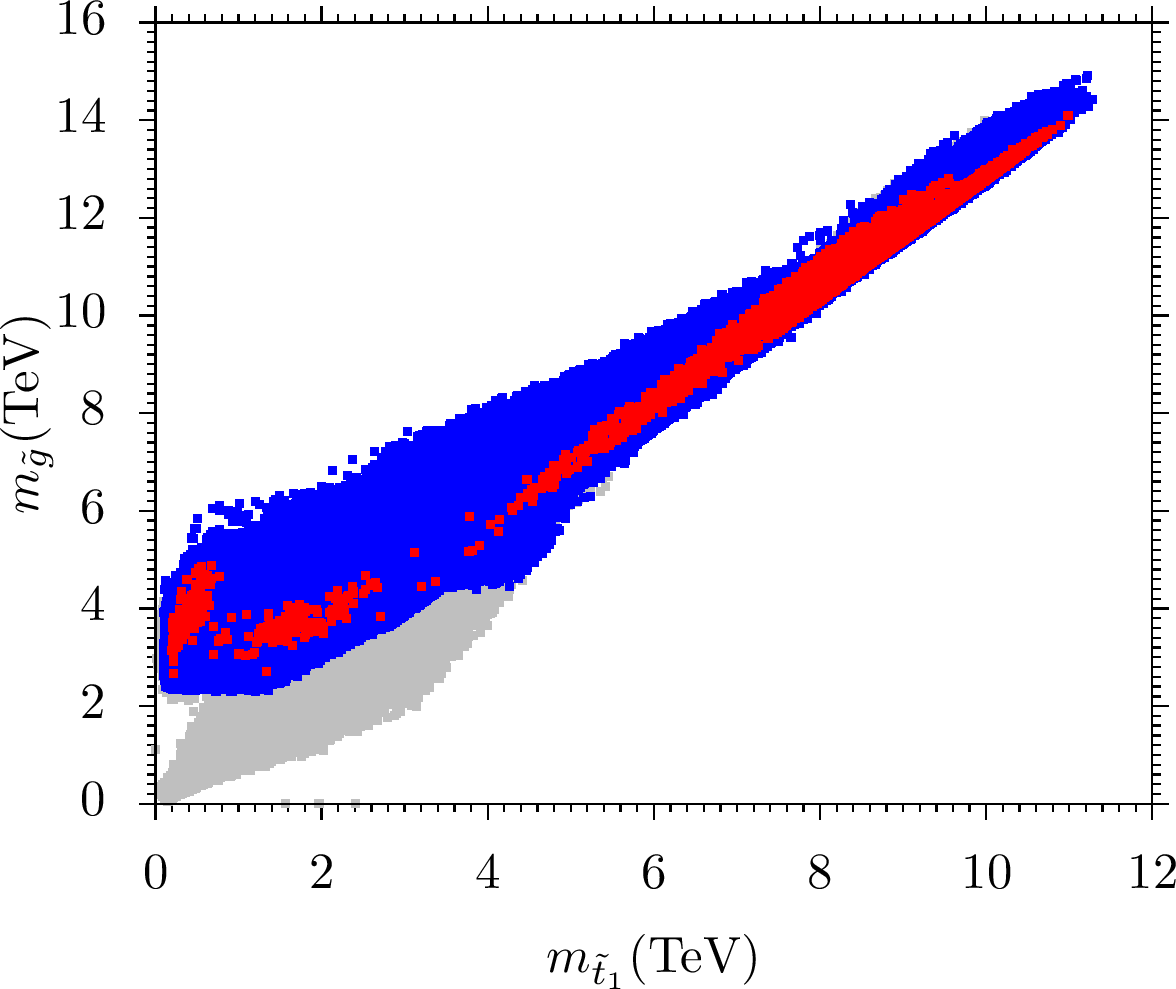}
	\centering \includegraphics[width=7.90cm]{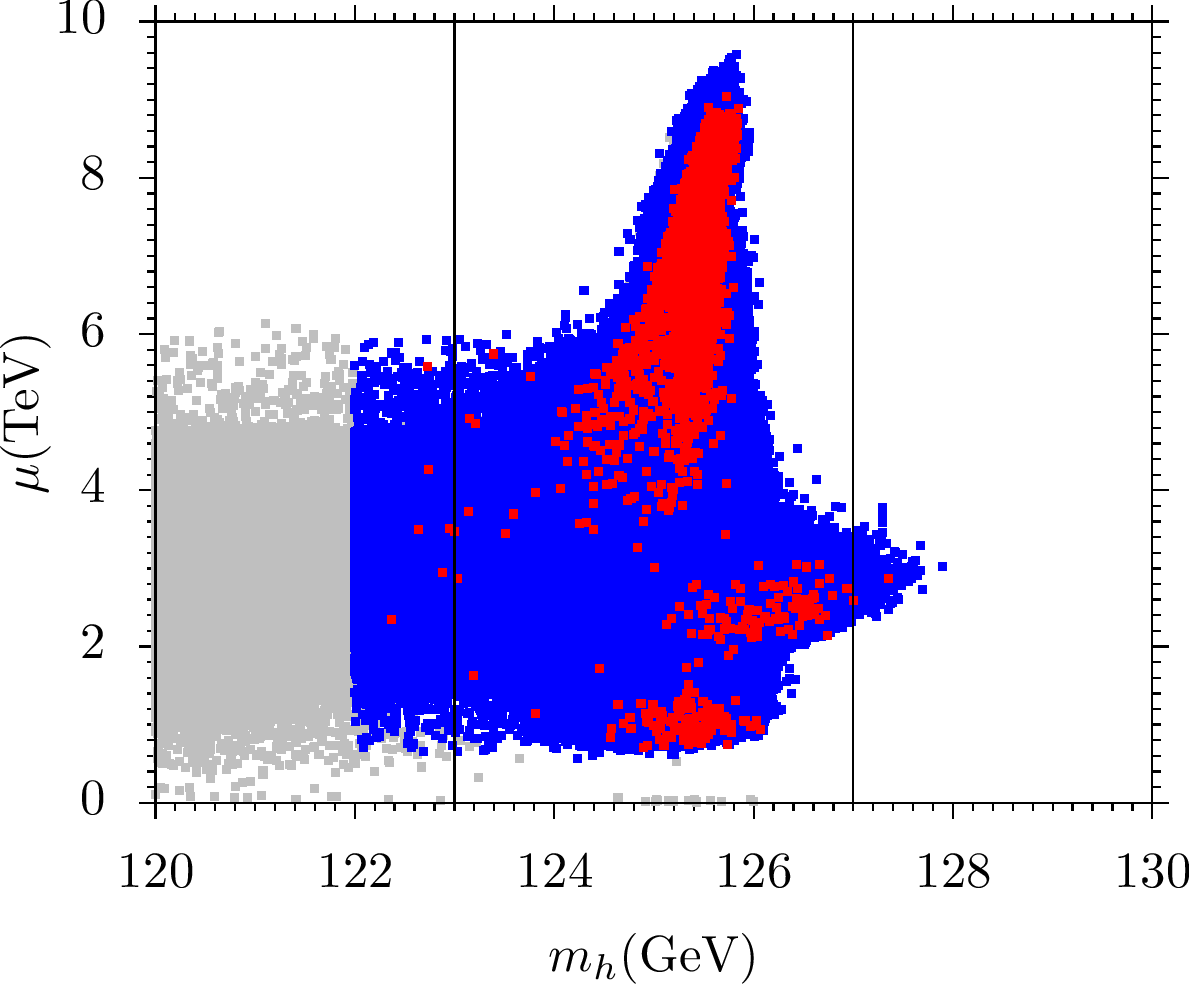}
	\caption{ Plots of results in  ${\widetilde{t}_1} - {m}_{\widehat{g}}$ and $m_h-\mu$ planes. The color coding and the panel description are the same as in Fig. (\ref{1}).} 
	\label{Delta_EW}
\end{figure} 
 
 In Fig.(\ref{Delta_EW}) we present ${m_{\widetilde{t}_1}} - {m}_{\widehat{g}}$ and $m_h-\mu$ planes. 
 The panel description and color coding are the same as in Fig.(\ref{1}). As we know LHC is a color particle machine and among the color sparticles, gluinos are the smoking guns for SUSY signals. As we have seen before we have heavy $M_{3}$ and also relatively heavy left-handed and right-handed scalars, consequentially we have heavy gluinos and stops. For both red and blue points, gluino mass is in the range of 2.2 TeV to 15 TeV, and stop mass $m_{\tilde t_{1}}$ is from 0.1 TeV to 11 TeV. It should be noted that at the 100 TeV $pp$ collider with 30$ab^{-1}$ integrated luminosity, gluino ($\tilde g$) mass can be probed up to 11 TeV and 17 TeV via heavy flavor decays and via light flavor decays respectively and stop ($\tilde t_{1}$) mass up to 11 TeV can be discovered \cite{Cohen:2013xda,Arkani-Hamed:2015vfh,Fan:2017rse,Golling:2016gvc}. In the right panel, we display the plot in the $m_h-\mu$ plane. Here we clearly see that both red and blue points are in the range [0.8,9] TeV. This implies that we have heavier higgsinos and fine-tuned solutions.

\begin{figure}[h!]
	\centering \includegraphics[width=7.90cm]{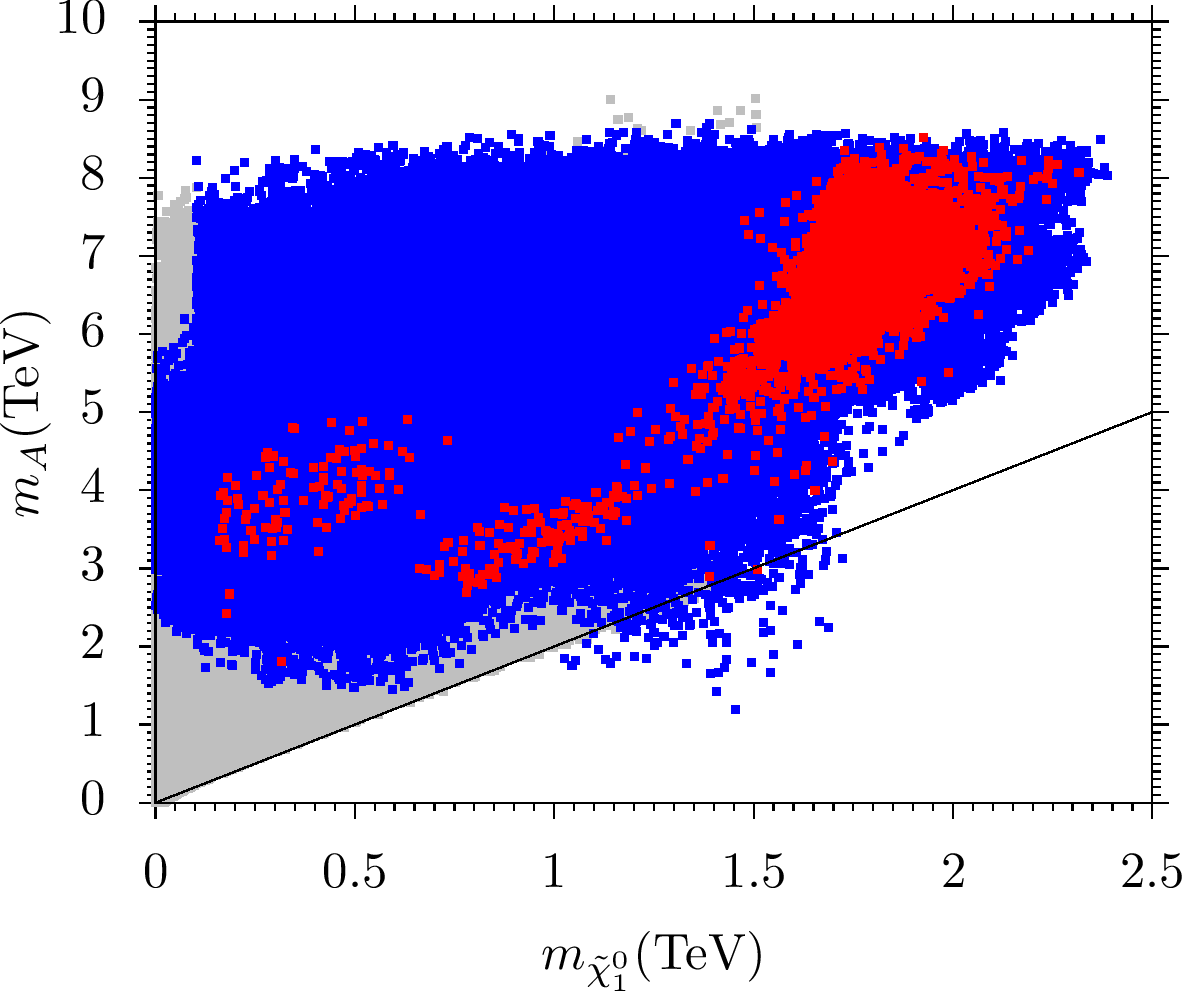}
	\centering \includegraphics[width=7.90cm]{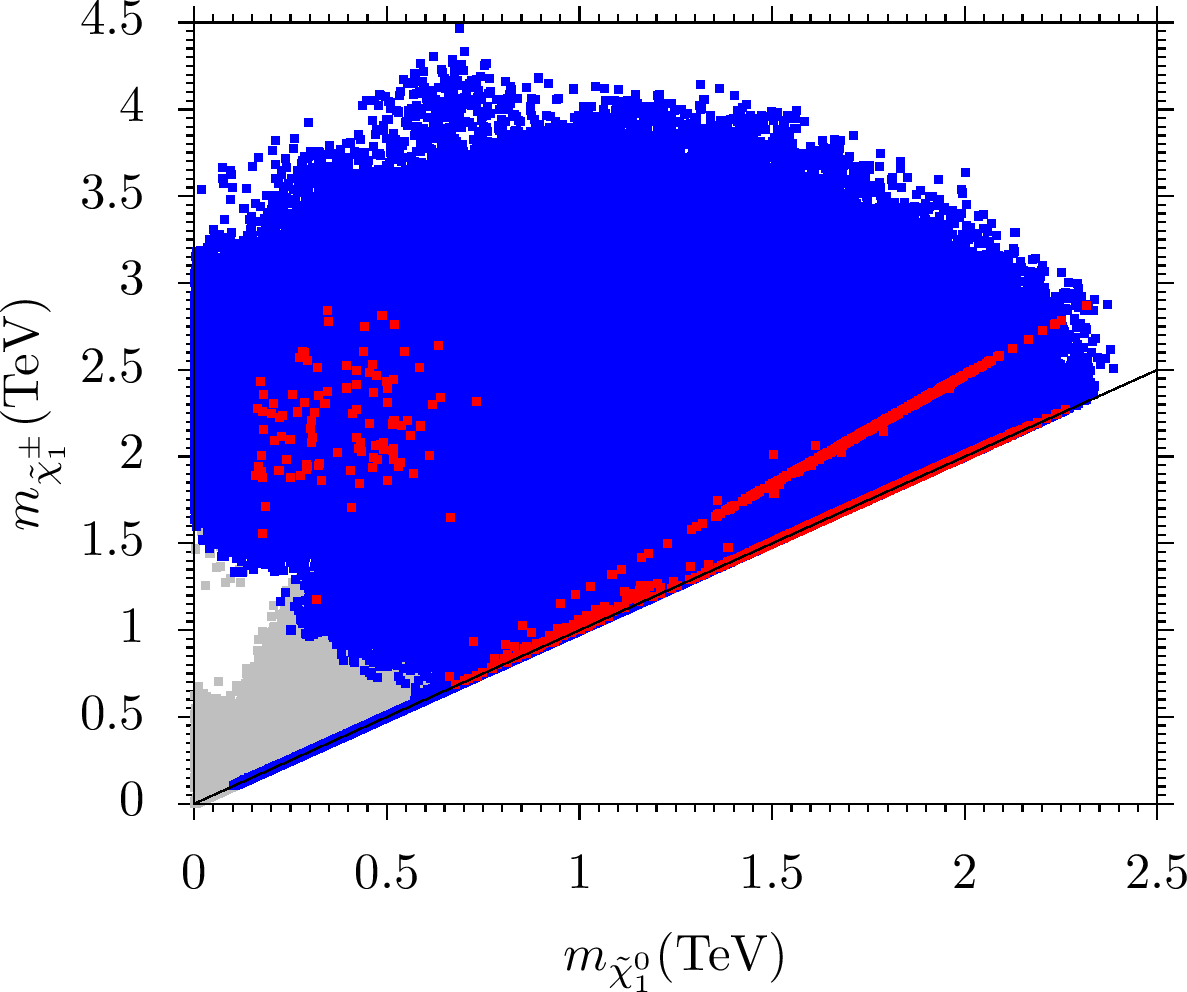}
	\centering \includegraphics[width=7.90cm]{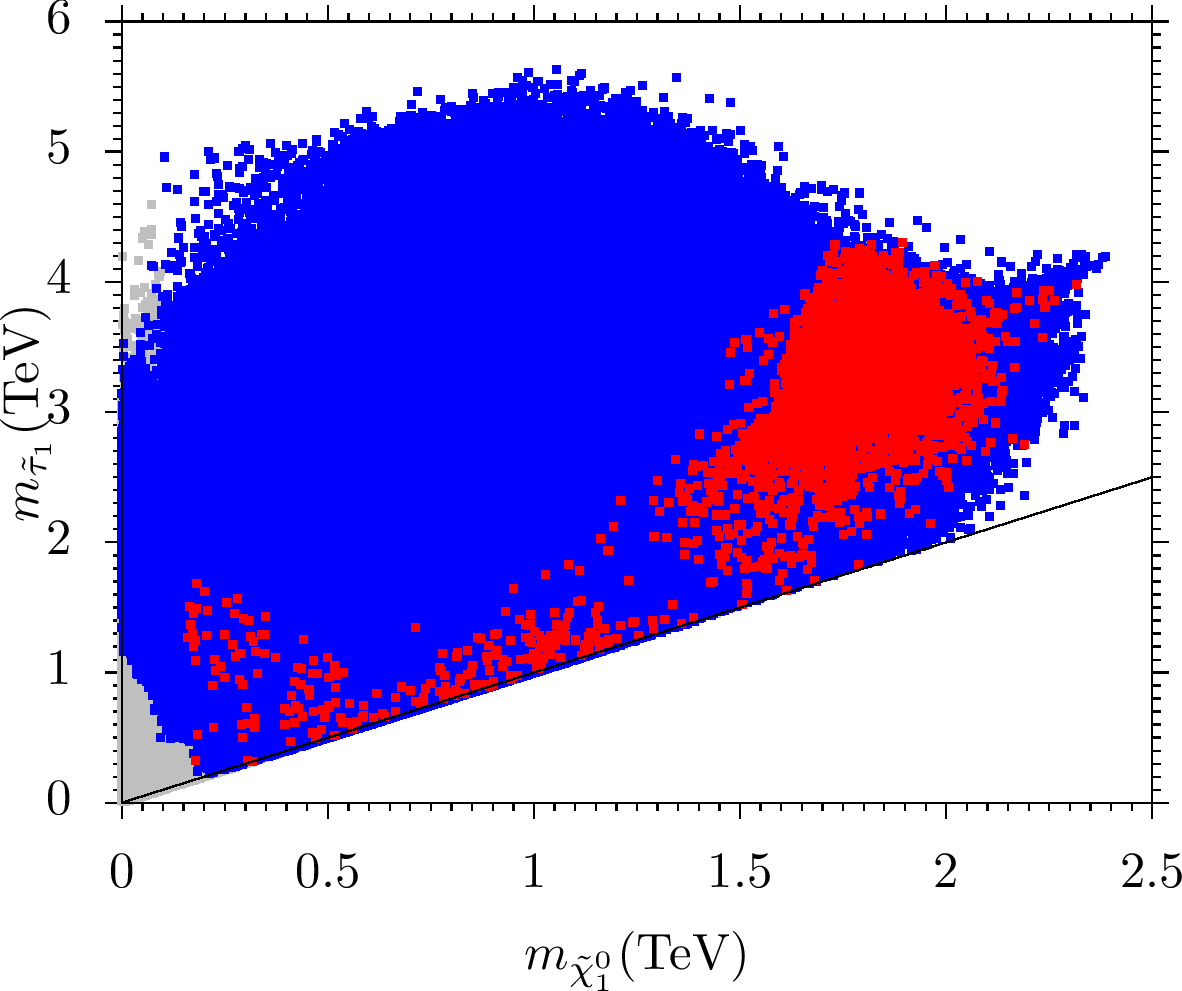}
	\centering \includegraphics[width=7.90cm]{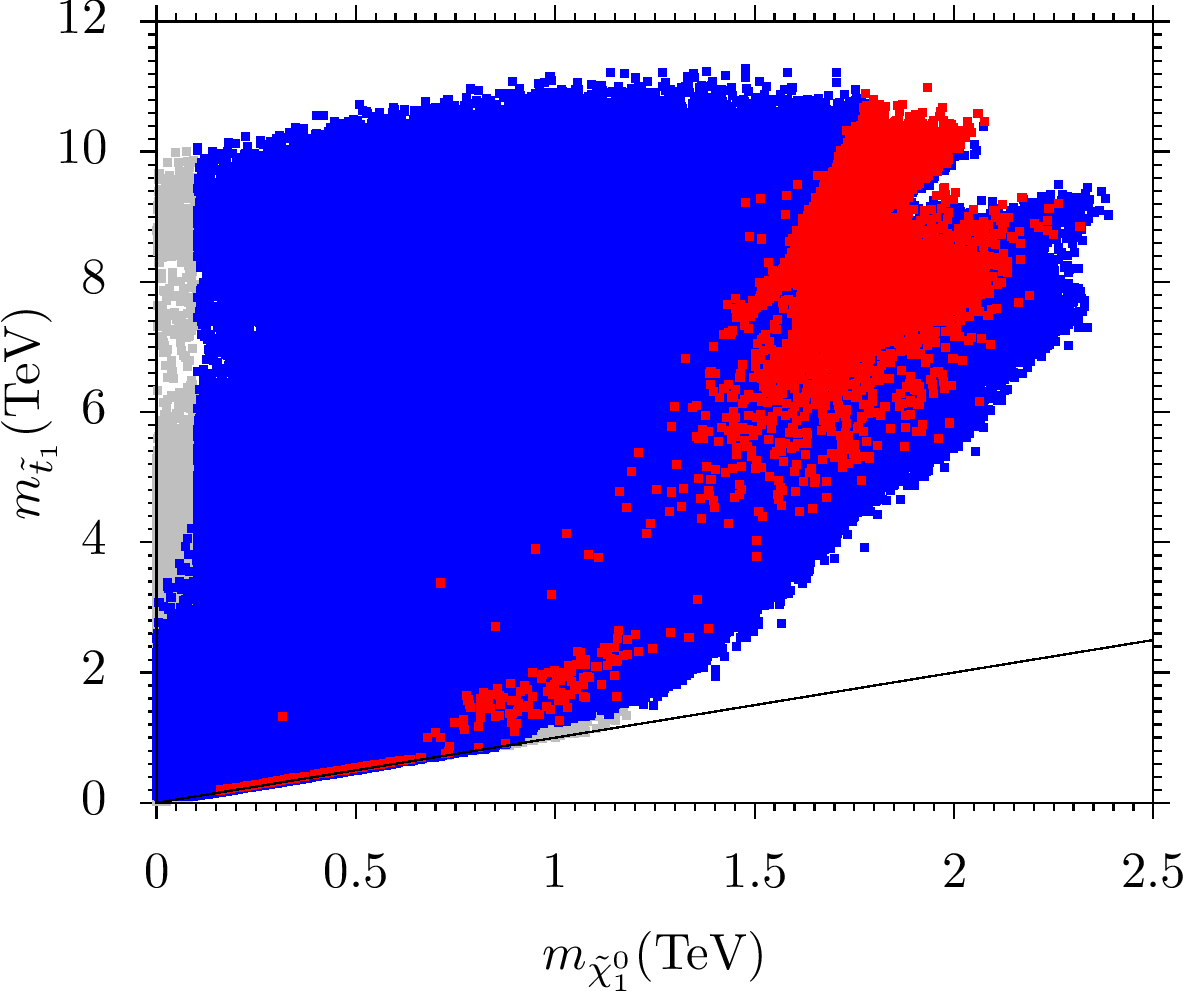}
	\caption{ Plots in $m_{\tilde{\chi}_1^0}$ - $m_A$, $m_{\tilde{\chi}_1^0}$ - $m_{\tilde{\chi}_1^{\pm}}$, $m_{\tilde{\chi}_1^0}$ - $m_{\tilde{\tau}_1}$ and $m_{\tilde{\chi}_1^0}$ - $m_{\tilde{t}_1}$ planes. Color coding and panel description are the same as in fig. (\ref{1})}.
	\label{4}
	\end{figure}
   
We now present results with the LSP neutralino mass and the masses of other particles of our model that are possibly light i.e. $\tilde \tau_1$, $A$, $\tilde \chi_{1}^{\pm}$, and $\tilde {t}_1$ masses in Fig. (\ref{4}). The description and color coding are the same as in Fig. (\ref{1}). The solid black lines show the mass degeneracy between the listed particles and for $m_{\tilde \chi_{1}^{0}}-m_A$ plane, indicates $m_A= 2 m_{\tilde{\chi}^0}$ region. In the top left panel, we present a plot in $m_{\tilde \chi_{1}^{0}}-m_A$ plane. We see that there are a couple of red points that have $m_A >{2 {\rm ~TeV}}$. In this scenario, correct relic density is achieved when a pair of LSP neutralinos annihilates into a CP-odd Higgs. It should be noted that for $m_A \lesssim$ 1.7 TeV is excluded for $\tan \beta \lesssim$ 30 \cite{CMS:2022goy}. In addition to it at Run 2, Run 3, and HL-LHC the CP-odd Higgs $A$ with $\tan\beta \lesssim$ 10 can be excluded for masses 1 TeV, 1.1 TeV, and 1.4 TeV respectively. We hope that future searches will be able to probe such solutions \cite{Baer:2022qqr,Baer:2022smj}.

In the top right panel, we show plot in $m_{\tilde \chi_{1}^{0}}-m_{\tilde \chi_{1}^{\pm}}$ plane. If we do not care about Planck2018 relic density bounds, we have neutralino and chargino degenerate masses solution from [0.1, 2.4] TeV but the degenerate masses solution is compatible with Planck2018 bounds from [0.7, 2.3] TeV range. The ref.\cite{ATLAS:2021ilc} has reported the 95$\%$ exclusion for sleptons as well as SM-boson mediated decays of $\tilde \chi_{1}^{+}\tilde \chi_{1}^{+}$ and $\tilde \chi_{1}^{\pm}\tilde \chi_{2}^{0}$. It can be seen that the charginos heavier than 300 GeV are safe when they are mass-degenerate with the LSP neutralino. On the other hand in the parameter space where slepton masses are heavier than charginos, these slepton-mediated decays will not take place. Since we also have heavier NLSP chargino solutions, we hope that such solutions will be probed in future LHC searches.  In the down left panel, we present the plot in $m_{\tilde {\chi}^0} - m_{\tilde{\tau}_1}$ plane. Here, if we observe that the range of red points where $\tilde{\tau}_1$ is nearly degenerate with $m_{\tilde {\chi}^0}$ is from [0.3, 1.8] TeV but for the blue points the mass degeneracy ranges are from [0.15, 2.1] TeV. Thus we note that our solutions are consistent with the results reported in \cite{CMS:2022rqk} with 137$\rm fb^{-1}$ at 13 TeV. 

 In the down left panel we present the plot in $m_{\tilde {\chi}^0}-m_{\tilde{t}_1}$ plane. Here we have red points $0.2 \, {\rm TeV} \lesssim m_{\tilde t_{1}} \lesssim  0.9 \, {\rm TeV}$ along the solid line. Such solutions represent scenario where NLSP stop is mass degenerate with the LSP neutralino. In such a case  $\tilde t_{1}\rightarrow c \tilde \chi_{1}^{0}$ is dominant decay channel. From the latest study \cite{ATLAS:2021kxv}, it is evident that in such a scenario the stop mass around 600 GeV has been excluded \cite{ATLAS:2023lon}. Thus nearly half of our solutions have already been excluded. We anticipate that the near future studies will probe the remaining NLSP stop solution in a small mass gap region. 

\begin{figure}[h!]
		\centering \includegraphics[width=7.90cm]{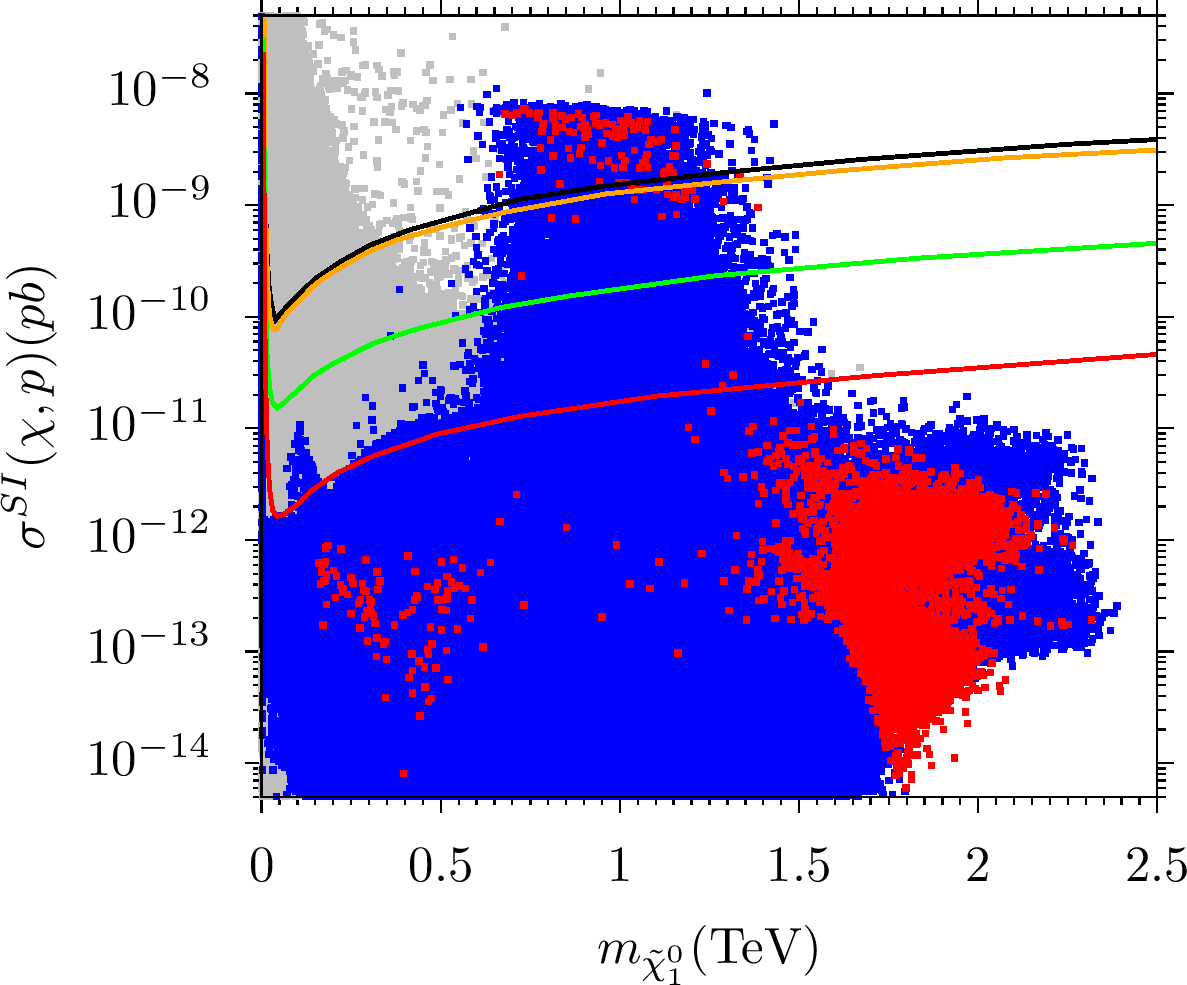}
		\centering \includegraphics[width=7.90cm]{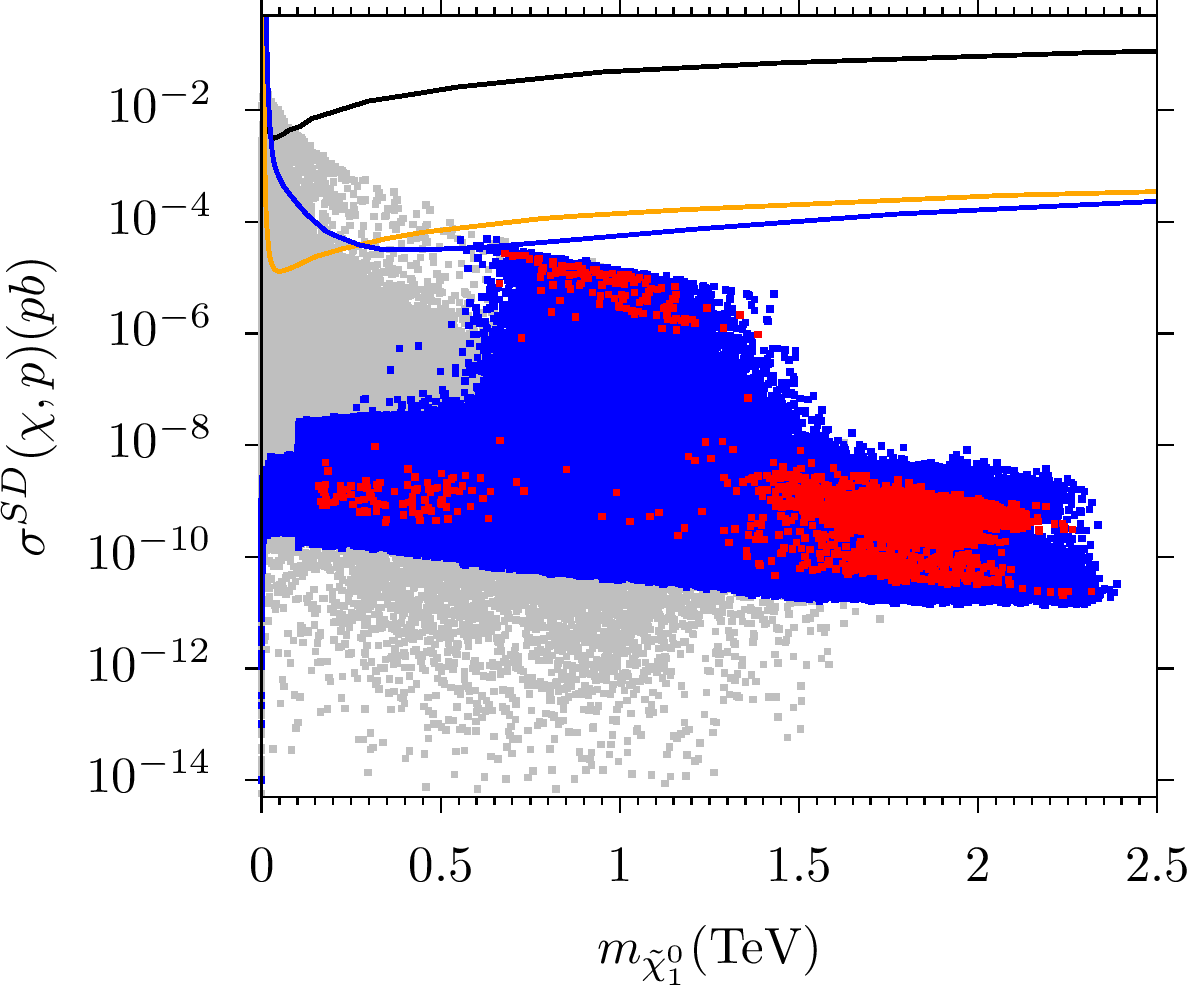}
		\caption{\small The spin-independent (left) and spin-dependent (right) neutralino-proton scattering cross-section vs. the neutralino mass. In the left panel, the solid black and orange lines depict the current LUX \cite{Akerib:2016vxi} and XENON1T \cite{Aprile:2017iyp} bounds, and the solid green and red lines show the projection of future limits \cite{Aprile:2015uzo} of XENON1T with 2 $t\cdot y$ exposure and XENONnT with 20 $t\cdot y$ exposure, respectively. In the right panel, the black solid line is the current LUX bound \cite{Akerib:2017kat}, the blue solid line represents the IceCube DeepCore \cite{IceCube:2009iyf}, and the orange line shows the future LZ bound \cite{Akerib:2016lao}. The color code in the description is the same as in the Fig. (\ref{1}).}
		\label{5}
	\end{figure}

  \begin{table}[h!]\hspace{1.0cm}
   \centering
  		\begin{tabular}{|c|cccc|}
  			\hline
  			\hline
  			& Point 1 & Point 2& point 3&Point 4\\
  			
  			\hline
  			$\widetilde{m}_{L}$        &1242.9& 1931.8&1779.7&905.2
  			\\
  			$\widetilde{m}_{R}$      &1748.8 & 787.7&4152&748.2 \\
  			$M_{1} $     &1630.2 & -1989.9&-2943.9&-1871.1  \\
  			$M_{2}$    &887.3  &-2120.4&-2256.8&-1713.6
  			\\
  			$M_{3}$ &2164.8& -1796.9&-4523.8&-1421.3\\
  			$A_0$ &-2475.7 & 4032.3&4920.8&3151.6
  			\\
  			$\tan\beta$   &16.3 & 12&39.5&16.9\\
  			$\widetilde{m}_{H_u}=\widetilde{m}_{H_d}$ &987 & 3481&1122.3&2634.9\\
  			\hline
  			$\mu$       &2944.2 &999.7&5797&859\\
  			\hline
  			
  			\hline
  			$m_h$           &122.8 & 125.7&125.4&125.1
  			\\
  			$m_H$           &3064  & 3804&2919&2842 \\
  			$m_A$            &3044  & 3779&$\boldsymbol{2900}$&2823\\
  			$m_{H^{\pm}}$   &3065  & 3805&2921&2843\\
  			
  			\hline
  			$m_{\tilde{\chi}^0_{1,2}}$
  			&$\boldsymbol{711}$, -724& $\boldsymbol{876}$, -1015 &$\boldsymbol{1388}$, 1694 &$\boldsymbol{807}$, -873 \\
  			
  			$m_{\tilde{\chi}^0_{3,4}}$
  			&2954, -2955   & -1015, 1763&-5786, 5787&897, 1413  \\
  			
  			$m_{\tilde{\chi}^{\pm}_{1,2}}$
  			& $\boldsymbol{724}$, 2958  &985, 1738&1698, 5783&838, 1398 \\
  			\hline
  			$m_{\tilde{g}}$ &4546 & 3810 & 8888 & 3049\\
  			\hline $m_{ \tilde{u}_{L,R}}$
  			&4096, 4284  & 3994, 3366 & 7874, 8552&2967, 2763   \\
  			$m_{\tilde{t}_{1,2}}$
  			&3376, 3693   & $\boldsymbol{911}$, 3237& 6428, 7095&1164, 2346  \\
  			\hline 
  			$m_{ \tilde{d}_{L,R}}$
  			&4097,4274  & 3343, 3995 &7875, 8536&2968, 2739 \\
  			$m_{\tilde{b}_{1,2}}$
  			&3656, 3693  & 3234, 3283&6496, 7416&2362, 2604 \\
  			\hline
  			$m_{\tilde{\nu}_{1,2}}$
  			&1380  & 2369&2559&1450\\
  			$m_{\tilde{\nu}_{3}}$
  			&1338 & 2336&2258&1383\\
  			\hline
  			$m_{ \tilde{e}_{L,R}}$
  			&1386, 1847  & 2370, 1064&2550, 4133&1455, 1017  \\
  			$m_{\tilde{\tau}_{1,2}}$
  			&1342, 1784  & 912, 2338 & 2554, 3732&$\boldsymbol{821}$, 1390  \\
  			\hline
  			
  			$\sigma_{SI}({\rm pb})$
  		&$2.56\times 10^{-12}$& $7.52\times 10^{-10} $& $2.88\times 10^{-13} $& $3.84\times 10^{-09} $\\
  			
  			$\sigma_{SD}({\rm pb})$
     &$2.17\times 10^{-9}$ & $1.98\times 10^{-06} $& $7.60\times 10^{-11} $& $1.1\times 10^{-05} $\\
  			
  			$\Omega_{CDM}h^{2}$ & 0.119 & 0.119 & 0.117&0.114\\
  			\hline
  			\hline
  		\end{tabular}
  		\caption{All quantities with mass dimension are in the unit of GeV and $\mu>0$. All points satisfy the particle mass bounds, B-physics constraints, and Planck bounds described in Section~\ref{constraints}. Point 1, represents chargino-neutralino coannihilation while point 2 neutralino-stop coannihilations. Point 3 depicts $A$-resonance; and finally, point 4 displays neutralino-stau coannihilation.
  		}
  		\label{table1}
  	\end{table}
  	

In Fig. (\ref{5}), we present the plots for spin-independent (left) and spin-dependent (right) neutralino-proton scattering cross section vs. the neutralino mass. We consider the impact of current and future dark matter (DM) searches on our model. 
In the left panel, the solid black and yellow lines respectively represent the current LUX \cite{Akerib:2016vxi} and XENON1T \cite{Aprile:2017iyp} bounds, whereas the green and red lines depict the projection of future limits \cite{Aprile:2015uzo} of XENON1T with 2 $t\cdot y$ exposure and XENONnT with 20 $t\cdot y$ exposure, respectively. In the right panel, the solid black line represents the current LUX bound \cite{Akerib:2017kat}, the orange line represents the future Lux-Zeplin (LZ) bound \cite{Akerib:2016lao} and the blue line represents the IceCube DeepCore.ref \cite{IceCube:2009iyf}. It can be seen that most points are consistent with current LUX and XENON1T bounds. Some points will also be probed by future Xenon experiments. On the other hand, there are some points that are excluded by the current LUX and Xenon1T experiments. Such points represent the scenario where chargino is the NLSP and the LSP neutralino is bino-higgsino mixed dark matter. We also want to make a comment here that our NLSP stop solutions are constrained by the collider searches below 600 GeV and the remaining solutions are constrained by the LUX and Xenon experiments. We note that points with NLSP mass around 900 GeV ( which is the heaviest NLSP stop in our model) have $\sigma^{SI}(\chi,p)$ just below the current LUX and Xenon1T bounds. Thus future DM searches will definitely prob such solutions.

In the right panel, the black solid line represents the current LUX bound \cite{Akerib:2017kat}, the orange line represents the future Lux-Zeplin (LZ) bound \cite{Akerib:2016lao} and the blue line represents the IceCube DeepCore \cite{IceCube:2009iyf}. Here we see that all solutions are consistent with current and future dark matter searches.

 To be concrete, we also present a table of benchmark points from our data which explain various scenarios of our discussion. In table-\ref{table1}, all points satisfy the constraints described in Section \ref{constraints}, and masses are given in GeV. Point 1 is an example of a chargino-neutralino coannihilation scenario. Here $m_{\tilde \chi^{0}_{1}}=$ 0.711 TeV and $m_{\tilde \chi^{\pm}_{1}}=$ 0.724 TeV. Point 2 shows stop-neutralino case where $m_{\tilde \chi^{0}_{1}}=$ 0.876 TeV and $m_{\tilde \chi^{\pm}_{1}}=$ 0.911 TeV. Point 3 represents $A/H$ resonance solutions with $m_{A}(m_{H})=$ 2900 GeV (2919 GeV). Finally point 3 displays stau-neutralino scenario with $m_{\tilde \chi^{0}_{1}}=$ 0.807 TeV and $m_{\tilde \chi^{\pm}_{1}}=$ 0.821 TeV. We also note except point 2 and point 4 have $\mu \lesssim$ 1 TeV which means these are relatively less fine-tuned solutions.

 \section{Summary and Conclusion}
 \label{summary}
 Because there are a few typos in the supersymmetry breaking sfermion masses and trilinear soft term, we revisit the phenomenological survey of the intersecting D-brane model with modified soft SUSY terms, focused on the LHC and DM constraints, and predict Low-energy SUSY particle spectra. The three-family Pati-Salam models have been constructed systematically in Type IIA string theory on the $\mathbf{T^6/(\Z_2\times \Z_2)}$ orientifold with intersecting D6-branes~\cite{Cvetic:2004ui}. Our phenomenological survey of this three-family Pati-Salam model has been presented in detail in Section \ref{Discussion}. In this work, we display the viable parameter space satisfying the collider and DM bounds along with the Higgs mass bounds.
We show that in our present scans, we have A/H-resonance solutions, chargino-neutralino coannihilation, stau-neutralino coannihilation, and stop-neutralino coannihilation. In the case of resonance solutions, $m_{A/H}$ is about 2 TeV or so.  In the case of chargino-neutralino coannihilation is concerned the NLSP chargino mass can be between 0.7 TeV to 2.3 TeV and the NLSP stau is in the mass range of 0.2 TeV to 1.8 TeV. As far as the NLSP stop solutions are concerned we we have solutions from 0.15 TeV to 0.9 TeV. Most of the parameter space related to this scenario has already been probed by the LHC SUSY searches. It should also be noted that the above-mentioned solutions, except for some of the chargino-neutralino solutions, are consistent with the ongoing and future astrophysical dark matter experiments. 

 \section*{Acknowledgement}
 TL is supported in part by the National Key Research and Development Program of China Grant No. 2020YFC2201504, by the Projects No. 11875062, No. 11947302, No. 12047503, and No. 12275333 supported by the National Natural Science Foundation of China, by the Key Research Program of the Chinese Academy of Sciences, Grant NO. XDPB15, by the Scientific Instrument Developing Project of the Chinese Academy of Sciences, Grant No. YJKYYQ20190049, and by the International Partnership Program of Chinese Academy of Sciences for Grand Challenges, Grant No. 112311KYSB20210012.


\end{document}